\documentclass[twoside,fleqn]{article}

\usepackage{espcrc2}
\usepackage{epsfig}
\newcommand{\twothird}{\mbox{\small $\frac{2}{3}$}}

\begin{document}
 

\title{\vspace{-3.65cm}
       {\normalsize DESY 04-168} \\[-0.2cm]
       {\normalsize LU-ITP 2004/028} \\[-0.2cm]
       {\normalsize HU-EP-04/58} \\[-0.2cm]
       {\normalsize Edinburgh 2004/22} \\[-0.2cm]
       {\normalsize LTH 635} \\
       \vspace{1.4cm} 
Axial and tensor charge of the nucleon with dynamical 
fermions\thanks{Talk presented by M. G\"ockeler at Lattice 2004.}}

\author{
A. Ali Khan\address{Institut f\"ur Physik, Humboldt Universit\"at zu Berlin,
                    D-10115 Berlin, Germany},
M. G\"ockeler\address{Institut f\"ur Theoretische Physik, 
                      Universit\"at Leipzig, D-04109 Leipzig, Germany}$^,$%
\address[R]{Institut f\"ur Theoretische Physik,
     Universit\"at Regensburg, D-93040 Regensburg, Germany},
P. H\"agler\address{Department of Physics and Astronomy, Vrije Universiteit,
   De Boelelaan 1081, 1081 HV Amsterdam, The Netherlands},
T.R. Hemmert\address{Physik-Department, Theoretische Physik,
          TU M\"unchen, D-85747 Garching, Germany},
R. Horsley\address{School of Physics, University of Edinburgh,
          Edinburgh EH9 3JZ, UK},
A.C. Irving\address[Liv]{Department of Mathematical Sciences,
          University of Liverpool, Liverpool L69 3BX, UK},
D. Pleiter\address[NIC]{John von Neumann-Institut f\"ur Computing NIC,
         Deutsches Elektronen-Synchrotron DESY, \\ D-15738 Zeuthen, Germany},
P.E.L. Rakow\addressmark[Liv],
A. Sch\"afer\addressmark[R],
G.~Schierholz\addressmark[NIC]$^,$%
\address{Deutsches Elektronen-Synchrotron DESY,
      D-22603 Hamburg, Germany}, 
H. St\"uben\address{Konrad-Zuse-Zentrum f\"ur Informationstechnik Berlin,
              D-14195 Berlin, Germany}
and J. Zanotti\addressmark[NIC] \\
QCDSF-UKQCD Collaboration}
 
\begin{abstract}
We present preliminary results for the axial and tensor charge of the 
nucleon obtained from simulations with $N_f=2$ clover fermions. A 
comparison with chiral perturbation theory is attempted.
\end{abstract}
 
\maketitle
 
Lattice QCD offers the possibility to investigate the internal structure 
of the nucleon without additional model assumptions. In particular, one
can study how the mass, the axial charge and further properties of the
nucleon change when the quark mass is varied or the nucleon is squeezed
into a finite spatial box. Even if we could simulate at the physical 
quark mass and in a huge volume, we would like to examine the dependence 
on the quark mass and on the volume, because it contains important 
information on low-energy QCD. This information can be extracted by
comparing the outcome of lattice QCD simulations with parameterisations
provided by chiral effective field theories (ChEFT). 

This comparison with ChEFT, if successful, leads to two
important achievements: Firstly, it gives us control over the chiral 
extrapolation and the thermodynamic limit of the simulation results 
and secondly it allows us 
to determine phenomenologically relevant coupling constants in ChEFT.

In the following, we shall present preliminary results from simulations with
$N_f = 2$ non-perturbatively improved Wilson quarks and the standard
plaquette action for the gauge fields obtained by the 
QCDSF and UKQCD~\cite{ukqcd} collaborations. 
Previous quenched QCDSF results can be found in Ref.~\cite{dis99} and for 
a review see Ref.~\cite{roger}. 
For the time being we do not yet perform an extrapolation to the 
continuum limit and neglect lattice artefacts.

Originally, ChEFT was invented to describe the quark mass dependence
of low-energy quantities by
means of effective hadron fields. However, for sufficiently large
volumes and sufficiently small pion masses, finite size effects are 
dominated by pions ``propagating around the world'', and therefore 
ChEFT can also be used to calculate the volume dependence: The effective
Lagrangian is independent of the volume. 

Our paradigm is the analysis of the nucleon mass~\cite{nmass}, where 
we could describe the pion mass dependence and the finite size effects
up to surprisingly large masses by formulae derived from relativistic
baryon chiral perturbation theory. We are trying to perform a similar
analysis also for the axial and the tensor charge of the nucleon, and
in the following we shall report on the status of these attempts.
Whenever we need numbers in physical units, we shall set the scale
using $r_0 = 0.5 \, \mbox{fm}$.

Compared with the case of the nucleon mass there are several additional
difficulties when dealing with such matrix elements of composite operators:
\begin{itemize}
\item
The ChEFT calculations are not as advanced as for the nucleon mass. 
\item
In order to suppress lattice artefacts, the operators should be 
improved. However, the required improvement coefficients are not yet 
known non-perturbatively. In the following, we shall present results 
from the unimproved operators.
\item
The operators have to be renormalised. We shall use non-perturbative
renormalisation factors computed by means of the Rome-Southampton
method with a linear chiral extrapolation at fixed bare gauge coupling.
\item
The quark-line disconnected contributions to the matrix elements are
hard to evaluate. Therefore we shall consider only flavour-nonsinglet
quantities, where these contributions cancel.
\end{itemize}

In order to compute the axial form factor of the nucleon $g_A (q^2)$ 
on the lattice we use the flavour-nonsinglet axial vector current
$ A_\mu^{u-d} =  \bar{u} \gamma_\mu \gamma_5 u - 
   \bar{d} \gamma_\mu \gamma_5 d $
and the form factor decomposition of its proton matrix element.
Using the so-called small scale expansion, which takes $\Delta$ degrees 
of freedom explicitly into account, the authors of Ref.~\cite{gasse} 
have studied the mass dependence of the nucleon axial charge 
$g_A \equiv g_A (q^2=0)$ to 
$O(\epsilon^3)$ and compared with quenched data.

\begin{figure}
\vspace*{-0.4cm}
\epsfig{file=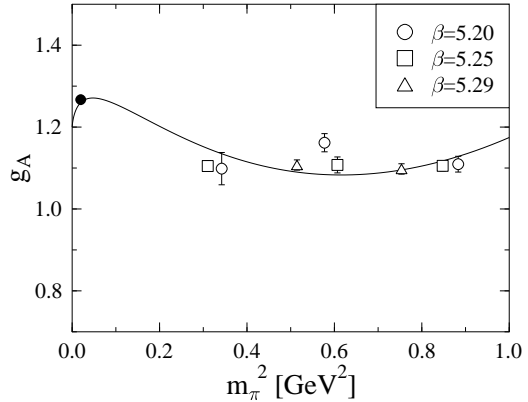,width=7.5cm} 
\vspace*{-1.0cm}
\caption{The axial charge $g_A$ of the nucleon. The filled circle indicates
the physical point.}
\label{fig.ga}
\end{figure}

In Fig.~\ref{fig.ga} we perform a similar comparison with our presently
available dynamical data for $g_A$ choosing some reasonable values for 
the parameters.
The curve reproduces
the weak mass dependence of the simulation results as well as the physical
point, at least approximately, but the chiral logarithm dominates 
only for very small pion masses.

\begin{figure}
\vspace*{-0.4cm}
\epsfig{file=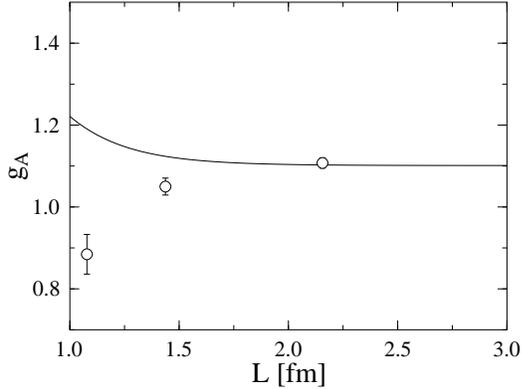,width=7.5cm} 
\vspace*{-1.0cm}
\caption{The volume dependence of $g_A$ for $m_\pi = 0.717 \, \mbox{GeV}$
         compared with leading order chiral perturbation theory.}
\label{fig.gafs}
\end{figure}

What about the finite size effects for $g_A$? There are indications that 
they could be quite large (for a quenched study see, e.g., 
Ref.~\cite{sasaki}), and they might not yet be negligible for 
the results in Fig.~\ref{fig.ga}. 
We have performed simulations on three different volumes at the same 
bare gauge coupling and quark mass, leading to $m_\pi = 0.717 \, \mbox{GeV}$ 
on the largest lattice, and found a clear volume dependence. 
First calculations within ChEFT have appeared~\cite{gafv}. To $O(p^3)$
in heavy baryon chiral perturbation theory (without explicit $\Delta$s) 
one finds 
\begin{displaymath} \begin{array}{l} \displaystyle
g_A(L) - g_A(\infty) = 
\frac{(g_A^0)^3 m_\pi^2}{6 \pi^2 f_\pi^2}
\sum_{\vec{n}} {}^\prime K_0(L |\vec{n}| m_\pi) 
\\[0.5cm] \displaystyle \quad {} 
 - \frac{g_A^0 m_\pi^2}{4 \pi^2 f_\pi^2}
\left( 1 + \twothird (g_A^0)^2 \right) 
\sum_{\vec{n}} {}^\prime \frac{K_1(L |\vec{n}| m_\pi)}{L |\vec{n}| m_\pi} \,,
\end{array}
\end{displaymath}
where $g_A^0$ is the value of $g_A$ in the chiral limit and $f_\pi$ is 
the pion decay constant. In 
Fig.~\ref{fig.gafs} we plot our data together with the above expression 
for the finite size effect. 
Not even the sign is reproduced correctly.
Of course, one may blame the large pion mass in connection 
with the low order of chiral perturbation theory for this discrepancy,
and further investigations are clearly needed.

\begin{figure}
\vspace*{-0.4cm}
\epsfig{file=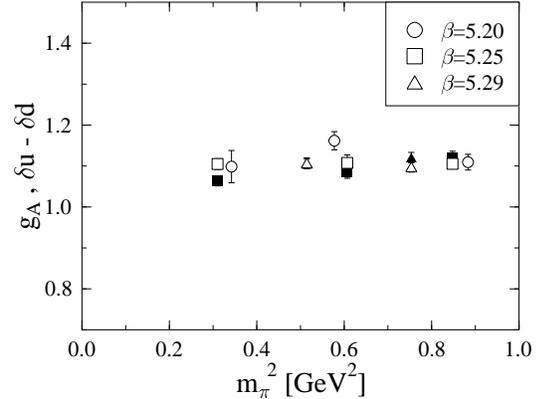,width=7.5cm} 
\vspace*{-1.0cm}
\caption{The tensor charge $\delta u - \delta d$ of the proton in the
         $\overline{\mbox{MS}}$ scheme at the renormalisation scale
         $\mu = 2 \, \mbox{GeV}$ (filled symbols) together with the 
         $g_A$ data (open symbols).}
\label{fig.combi}
\end{figure}

The tensor charge of the nucleon is the lowest moment of the 
transversity distribution $h_1$. For flavour $q$ it is given by the
proton matrix element
\begin{displaymath}
\langle p,s | \bar{q} \mathrm i \sigma_{\mu \nu} \gamma_5 q | p,s \rangle
 = \frac{2}{m_N} (s_\mu p_\nu - s_\nu p_\mu) \delta q
\end{displaymath}
with $s^2 = - m_N^2$. Again, we restrict ourselves to the 
flavour nonsinglet combination $\delta u - \delta d$ and plot it in 
Fig.~\ref{fig.combi} together with $g_A$ data, because in the non-relativistic 
limit one expects that both quantities agree. Hence this comparison gives
us some information on how relativistic the quarks in our simulations are
and the data seem to indicate that the quarks are not very relativistic.

Concluding we may say that the interpretation of our results is only at 
its beginning. Once we have analysed our configurations completely, we 
hope to gain more insight into the structure of the nucleon. But further
improvements are certainly desirable, in particular smaller quark masses
and lattice spacings in the Monte Carlo simulations as well as more 
extensive calculations in ChEFT.

\section*{ACKNOWLEDGEMENTS}

The numerical calculations have been performed on the Hitachi SR8000 at
LRZ (Munich), 
on the Cray T3E at NIC (J\"ulich) and ZIB
(Berlin), as well as on the APE1000 and Quadrics at DESY (Zeuthen). We
thank all institutions for their support.
This work has been supported in part by 
the EU Integrated Infrastructure Initiative Hadron Physics (I3HP)
and by the DFG (Forschergruppe Gitter-Hadronen-Ph\"anomenologie). 
A. A K acknowledges support by ``Berliner Programm zur F\"orderung der 
Chancengleichheit f\"ur Frauen in Forschung und Lehre''.

\end{document}